# Novel therapeutic targets in Chronic Myeloid Leukaemia through a discrete time discrete Markov chain model of BCR-ABL1 interactions.

Alfonso Vivanco-Lira[1,2].

[1]Undergraduate medical student.

Medical Sciences Department.

University of Guanajuato.

Leon, Guanajuato, Mexico.

[2]Undergraduate mathematics student.

Exact Sciences and Engineering Division.

Open and Distance Learning University of Mexico.

Mexico City, Mexico.

Email: poncho9715@gmail.com

Summary.

Chronic Myeloid Leukaemia (CML) is a blood-derived proliferative disorder, which is highly associated to a translocation of chromosomes 9 and 22 or the creation of Philadelphia chromosome Ph(+) cases, inducing the synthesis of a chimeric fusion protein, namely BCR-ABL1 (Breakpoint Cluster Region-Abelson 1 chimeric protein), which is known for driving the pathophysiology of the disease, however variants of CML are also recognized as CML Ph(-), these nonetheless account for a small percentage of the overall CML patients; posing thus the question whether BCR-ABL1 fusion protein is required for the whole of the pathophysiology of CML. Hereof, through a stochastic description, a discrete time discrete Markov chain depicts the various protein-protein interactions of BCR-ABL1 to better understand signalling pathways and time-dependent evolution of these pathways, as well as to provide prospective therapeutic protein targets to improve both the specificity of the treatment and the life-expectancy of the patients.

Background.

*Definition.*

Chronic myeloid leukaemia (CML) is a blood cancer in which there is an abnormal proliferation of myeloid-derived cells, such as neutrophils, basophils, eosinophils, mast cells, monocytes; this pathology is classically associated to a translocation in the chromosomes 9 and 22 t(9;22) showing afterwards the expression of a fusion protein, BCR-ABL protein, ABL1 from the Abelson protein portion on chromosome 9 and BCR from the breakpoint cluster region on chromosome 22 (Jabbour & Kantarijan, 2018), BCR-ABL-negative cases of chronic myeloid leukaemia have been reported and account for around 1 to 2% of the CML cases, in which there exist various genetic aberrations, such as: mutations in CSF3R, SETBP1, JAK2V617F, KRAS, NRAS (Giri, Pathak, Martin, & Bhatt, 2015), although this could be disputed due to the fact that these patients show genetic instability thus, the translocation could vary in time (Balk, Fabarius, & Haferlach, 2016).

*Epidemiology.*

The incidence of CML varies from 4 per million people in non-Western countries to 1.75 per 100,000 people in the USA, this variation could be due the fact that BCR-ABL negative clones are sometimes not taken in account as CML cases. The incidence shows an increase with corresponding increasing age, for it has been shown that in some European countries the median age of diagnosis is from 57-60 years, nevertheless cases have been seen from 15 years to more than 85 years of age, with two peaks of age at diagnosis in the intervals from 70 to 74 years and from 80 to 84 years.

The prevalence has been reported from 5.6 to 7.3 per 100,000 inhabitants (Höglund, Sandin, & Simonsson, 2015), and estimations have been made in order to predict the prevalence in upcoming years which show a prevalence of 140,000 patients in 2030 and of 180,000 patients in 2050 in the USA alone, showing then an increase in the prevalence (Huang, Cortes, & Kantarijan, 2012).

Risk factors have been associated to CML, such as the exposure to ionising radiation as it has been seen in people who survived the atomic bomb in Hiroshima (Hehlmann, Hochhaus, & Baccarani, 2007) and in people in whom it has been performed both radiography and tomography in repeated occasions (Ju, et al., 2016). In regards to chemicals, a meta-analysis showed that there exists no relationship between benzene and CML development, with a $OR = 1.003$ (Lamm, Engel, Joshi, Byrd III, & Chen, 2009). Another study showed the potential relationship in a dose-dependent manner between cigarette smoking and CML development, with a OR increasing with the increasing amount of pack-years, however there is no overall relationship between the smokers vs. non-smokers group in CML development (Qin, Deng, Chen, & Wei, 2016). The overall risk of developing CML is greater in males than in females with a male-to-female ratio which varies from 1.2 to 1.7 (Höglund, Sandin, & Simonsson, 2015).

*Pathophysiology.*

As in the previous remarks it has been seen, one of the main features of CML is the phenomenon of translocation of the chromosomes 9 and 22, i.e., t(9;22)(q34;q11) then adjoining two distinct genes, the BCR gene (breakpoint cluster region) and ABL1 gene (Abelson), this juxtaposition occurs most of the occasions at the intro 13 or 14 of BCR and the exons 1b and 2 of ABL1 (Chereda & Melo, 2016) however different juxtaposition regions have been observed, such as in the intron 14 of BCR and the intron 2 of ABL1 (Lyu, et al., 2016) and many others: b3a2, b2a2, b3a3, b2a3, e1a2, 19a2 (Melo, 1996).

The BCR gene is located at 22q11.23, and shows 23 exons and it is expressed widely throughout the body, it has been associated to both acute as well as to chronic myeloid leukemia (NIH, 2019), this BCR gene encodes two RNA species of 4.5 and 6.7 kbp, and its promoter region contains SP1 binding sites, it has been proposed to have a role in metabolism; the translated protein product is a phosphoprotein with serine/threonine kinase activity as well as phosphotransferase activity, this activity has been confirmed with the homologies it shows in regards to other kinases (Maru & Witte, 1991). Some of its molecular interactions are shown in Table 1 (TyersLab, 2019), and we may postulate six sets of types of molecules with which it interacts: signal processing (P), cell trafficking (T), cell growth (G), cell proliferation (Pr), metabolism (M), structure (S) however these sets show obvious overlap, $P \cap M \neq \emptyset \wedge S \cap T \neq \emptyset \wedge P \cap Pr \neq \emptyset$ (Table 1).

ABL1 in turn is a protooncogene with tyrosine kinase activity which is involved in several cellular processes: cell division, adhesion, differentiation, response to stress. The activity of ABL1 protein is negatively regulated by its SH3 domain. It is displayed in various types of leukaemias. Possesses 12 exons (NIH, 2019).

The resulting fusion protein BCR-ABL1 is a 210 kDa protein, and sometimes it could be detected in healthy individuals hypothesizing the lack of potency in the transformed cell to support leukaemia (Ismail, Naffa, Yousef, & Ghanim, 2014). The domains in BCR-ABL1 which appear to affect the protein function are the Ser/Thr kinase domain pertaining to BCR, SH1 tyrosine kinase domain and the coil-coil protein dimerization domain. This fusion protein lacks the N-terminal myristoylated region of ABL1 which causes autoinhibition of the SH1 kinase activity. Thus, should we inhibit the myristoylated domain-binding pocket we would be doing so in an effective manner, it has been done through GNF2 and it has been seen to reduce the emergence of resistant mutants (Zhang, et al., 2010). BCR-ABL1 fusion protein promotes aberrances in apoptosis, proliferation and cell adhesion (through signalling pathways such as: JAK/STAT, PI3K/AKT, Ras/MEK, p53, MYC, beta-catenin, C/EBP$\alpha$), augmenting the proliferation of granulocytes and causing the clinical features seen in the chronic phase of CML (Chereda & Melo, 2016). The transition in time towards a blastic phase following an accelerated phase, has been proposed to be initiated through BCR-ABL1 genetic damage, and inadequate DNA repair machinery (Perrotti, Jamieson, Goldman, & Skorski, 2010).

*Treatment.*

We may in first order define what a complete cytogenetic response (CCR) is, which is the absence of the Ph in $\geq 20$ cells in metaphase in a bone marrow aspirate, and a major molecular response (MMR) is reached when $r = \frac{BCR-ABL1}{ABL1}$ (where BCR-ABL1 stands for the concentration of the chimeric protein) is $r < 1x10^{-3}$ or $-\log(r) > 3$ (Oriana, et al., 2013). Having these definitions as a preludium, we may say that imatinib mesylate stands as the gamechanger in CML and the course of the disease, with a recommended dose of 400 mg/day, however different doses have been proposed based upon both the MMR and CCR reached with each, the average CCR with imatinib at 400 mg is 57.67% and with 800 mg of 65.67%, while the MMR with 400 mg is 36.67% and with 800 mg is 45% (Aladag & Haznedaroglu, 2019); imatinib is a tyrosine kinase inhibitor which binds to the ABL1 portion of the chimeric protein, imatinib binds and stabilises the non-ATP binding conformation of BCR-ABL1 kinase domain. Imatinib has been combined in some fashions in order to promote synergy of drugs, these other compounds are arsenic, interferon-alpha, cytarabine, doxorubicin, leptomycin B, with some enhancements of imatinib properties. Resistances to imatinib have been seen and should be early characterized in order to escalate doses or to move to second line treatments (Moen, McKeage, Plosker, & Siddiqui, 2007).

Second line treatments are dasatinib and nilotinib, the latter exhibits better CCR and MMR (CCR=87%, MMR=71% in 24 months) as well as dasatinib (CCR=86%, MMR=82%. A combination treatment of nilotinib and Peg-IFN (interferon used to be the frontline treatment of CML before the arrival of imatinib) has been proposed and shows good results in 12 months with an MMR=76% and CCR=100%. As third-line treatments, there exist some options, such as the allogenic stem cell transplantation, and tyrosine kinase inhibitors: bosutinib and ponatinib (Aladag & Haznedaroglu, 2019).

Several combinations for tyrosine kinase inhibitors (TKI) have been proposed, such as: bosutinib or dasatinib or omatinib or nilotinib + IFN-alpha; imatinib + hydroxyurea/cytarabine/omacetaxine/homoharringtonine; ascimimib, immune modulation (nivolumab + dasatinib, avelumab + TKI), PPARgamma agonists, DPPIV inhibitors (Westerweel, te Boekhorst, Levin, & Cornelissen, 2019), other molecules that have been used in order to inhibit CML are: arsenic trioxide (as monotherapy), hydroxychloroquine, BMS-833923, LDE225, panobinostat, ruxolitinib, zileuton (Bhatia, 2017).

Model.

Hereby one model concerning the stochastic nature of the proteins associated with the fusion protein BCR-ABL1 in CML is displayed.

*Stochastic model.*

A stochastic model concerning the protein-protein interactions in CML could provides us with new ground and opportunities in therapeutic approaches in the short- and long-term future, through estimations of the most probable nodes or proteins in the network could we direct our efforts in order to develop new and more potent drugs, as well, through estimations of mean time of passage we can predict to which node our network tends to. This model may as well bring light upon the pathophysiology of the transition towards a blastic crisis. We will consider first the BCR protein network alone, followed by the one related to the ABL1 protein, and finally we will deem them as a joint set; In the thrice of cases, the models are to be discrete time discrete Markov chains.

The BCR protein has shown interactions with 87 different proteins (TyersLab, 2019), these interactions are then a stochastic process with a set of states $\hat{S} = \{P_1, P_2, \ldots, P_n\}$ where P stands for protein and then we can define a function f, such that $f: \hat{S} \to S_1$ where $S_1 = \{s_i \in \mathbb{Z}^+\}$ is a bijective function for the states of proteins. The temporal parameter is defined as $T = \{t_i \in \mathbb{Z}^+\}$ where this temporal parameter for the meantime shall be an arbitrary time unit. The rules of transition are,

$$s_i R s_i \to s_i$$
$$s_i R s_j \to s_j$$
$$s_j R s_i \to s_i$$

Where R stands for the presence of a relationship or interaction between de i-th and j-th proteins.

The stochastic matrix of this protein has the following shape,

$$P = \begin{pmatrix} p_{11} & p_{12} & p_{13} & \cdots & p_{1n} \\ p_{21} & p_{22} & p_{23} & \cdots & p_{2n} \\ p_{31} & p_{32} & p_{33} & \cdots & p_{3n} \\ \vdots & \vdots & \vdots & \ddots & \vdots \\ p_{n1} & p_{n2} & p_{n3} & \cdots & p_{nn} \end{pmatrix}$$

The complete matrix is shown (Fig. 1).

If we set the initial vector to be at equilibrium, i.e., $\pi^0 = \frac{1}{86}(1_1, \ldots, 1_{86})$, we may determine the distribution at any given moment $t_n$, according to the following expression,

$$\pi^n = \pi^0 P^n$$

First, let us find some distributions at selected values of n (Fig. 2). From this we can predict what the limit distribution would look like, and we can organize the proteins by their limit distribution from greater to lesser, here we will present the proteins in which the process will tend to be 95.2973% of the time from there will we be able to determine possibilities of directed therapy, and we will represent the information as logic circuits, the possible therapies $\forall a_i \in A \exists b_i \in B | B = \{b_i \in \mathbb{Z}^+\} \wedge f: A \to B$ where A stands for the set of the drugs available.

The matrix for ABL1 protein is as shown (Fig. 3). We will determine the distributions when $n \to \infty$ for ABL1 (Fig. 4).

This model provides us with the opportunity to observe the behaviour of the signalling networks in CML, more specifically, the probability of every network to be at any given state at one moment, interestingly do we notice that neither of both nodal proteins (BCR nor ABL1) show a large stationary probability, this invites us to think that therapeutic targets in CML is made up from a larger set of molecules than we may have initially considered, and that we may be able to deliver a much more complete therapy to CML patients in such a way that the backbone treatment is one corresponding to those interacting with either BCR or ABL1, but we then have the limbs of the treatment, where we can by all means induce a thrust by considering the probability of finding any of both systems at any state (of the most probable ones). In order to compute feasible "limb"-like treatments, we shall consider the drug efficacy of the molecules we will introduce, this in turn will allow us to compute the complete or total efficacy of the treatment. This feasibility must be taken in account from the stationary probability obtained above, from which we shall consider those molecules whose summed up probability equals 0.95, such that,

$$\sum_{i=1}^{n} \pi_i \geq 0.95$$

Where $\pi_i$ stands for the stationary probability for the system to be in the i-th state, this expression let's us find in which state the systems will be in 95% of the time. For the case of BCR protein, this means the available drugs for 63 different proteins (including BCR) and for ABL1 this means the equally available drugs for 133 different proteins, the first set of drugs let us define it as A and the second one as B,

$$|A \cap B| > |\emptyset|$$

Moreover, let us define a set C which is the set of all proteins with which BCR interacts and D as the set of all proteins with which ABL1 interacts,

$$|C \cap D| > |\emptyset|$$
$$\Omega_1 = \{C, D\}$$
$$\mathrm{p}(\Omega_1) = 1$$
$$C \subset D$$
$$D \subset C$$
$$\therefore C \neq D$$

Currently, the previously described sets, say A and B show the following relationships,

$$A \subset B$$
$$B \subset A$$
$$\therefore A \neq B$$

We define these sets,

$$A \subseteq C$$
$$A = \left\{ \forall a_i \in A | \sum_{i=1}^{n} \pi_i \geq 0.95 \right\}$$
$$B \subseteq D$$
$$B = \left\{ \forall b_j \in B | \sum_{j=1}^{m} \pi_j \geq 0.95 \right\}$$

Now, we may describe the following probability tree,

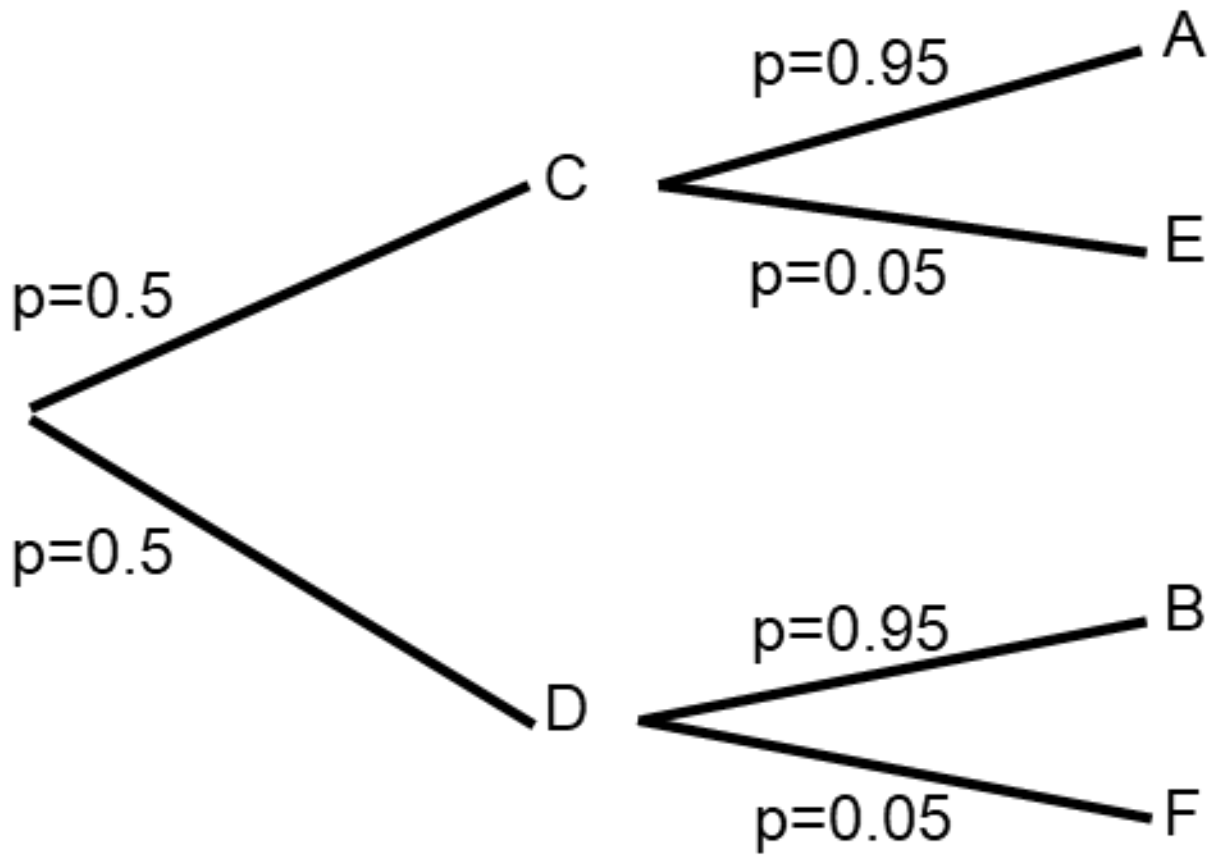

Thus, they do share proteins, more specifically,

$$|A \cap B| = 167$$

If we define a new probability space, $\Omega_2 = \{A, B\}$ with $p(\Omega_2) = 1$, then, this space defines those proteins that interact with both BCR and ABL1 proteins: the protein interactions corresponding to those of the chimeric protein. We can outline a new Markov chain from this probability space, such that the set of states $S_2 \subseteq S_1$, this set of states is the sum of the probabilities in $\Omega_1$ is $\sum_{i=1}^{n} \pi_i = 0.95$, nevertheless in $\Omega_2$ this sum results $\sum_{i=1}^{n} \pi_i = 1$, these states display stationary probabilities and are shown in Fig. 5.

Discussion.

Undoubtedly is imatinib the frontline and central treatment in CML, it indeed has changed the prognosis landscape for these patients, showing very good results in regards to the MMR and CCR, achieving 57.67% in CCR with 400 mg/day and an MMR with 400 mg/day of 36.67% (Aladag & Haznedaroglu, 2019) with a life expectancy of patients with CML of 3 years less than that of the general population and this life expectancy has been computed to be always less than that of the general population; moreover, the use of imatinib and other TKIs poses a risk predisposing to other cancers and cardiovascular morbidity (Bower, et al., 2016). Imatinib as the medullar treatment of CML, and other tyrosine kinase inhibitors display as their mechanism of action the binding to the chimeric protein BCR-ABL1, whether to the ABL1 portion or to the BCR portion of this protein and while they have approximated the life expectancy of patients with CML to that of the general population and improved their quality of life, efforts can be made to provide alternative therapeutic approaches to CML.

Through the model previously introduced, we have computed the stationary stochastic probabilities of three Markov chains, the ones corresponding to the BCR protein, ABL1 protein and the chimeric BCR-ABL1 protein; we then selected those proteins whose sum of stationary probabilities summed $p = 0.95$, then from those proteins, we computed a new chain, the one for the chimeric protein and its stationary probabilities. This showed that the network of proteins with which BCR-ABL1 interacts is a vast array composed of 167 proteins, including themselves, and that this network in the stationary spectrum or when the number of steps in time taken when $n \to \infty$ does favour the rationale of use of TKIs due to the fact that the stationary probabilities for BCR and ABL1 in the chimeric protein Markov chain are,

$$\pi_{ABL1} = 0.07784161$$
$$\pi_{BCR} = 0.04530407$$

Because these two probabilities are the maxima of the protein array,

$$\pi_{ABL1} > \pi_{BCR} > \pi_j$$

That is, they are greater than the stationary probability of any other j-th protein in the Markov chain. This model provides us with information about how much time the network spends on each of the members of the network based upon the interactions they show, thus they spend the $\pi_{BCR} + \pi_{ABL1} = 0.12314568$ of time ($t = 12.31457\%$) in these proteins, but this time is less than the eighth of the time of the network, and if good results are achieved through the inhibition of these chimeric protein (or the two states of the Markov chain), then the inhibition (or corresponding activation) of interacting proteins could improve the results in patients with CML; some proteins that show promises in this regard are those that interact both with the BCR part and the ABL1 part of the chimeric protein, $BCR \cap ABL1 = K$, where K is the set that contains these proteins: TP53, CBL, GRB2, HSPA8, PIK3R1, UBC, SHC1, STUB1, HSP90AA1, CRK, HSPA4, CRKL, RB1, PIK3R2, KIAA1429, TRIM25, UBASH3B, NTRK1, SOS1, DOK1, XPO1, AP2M1, PTPN6, HSPD1, RAD51, HCK, LRRK1 (with their corresponding stationary probabilities shown in Fig. 6) this subnetwork or set of proteins contains the 43.71% of the time spent in the network. This approach could be improved by observing the signalling pathways in which BCR-ABL1 engages and as we have previously mentioned, the main pathways involved in the pathophysiology of CML through BCR-ABL1 are: beta-catenin, C/EBP-alpha, JAK/STAT, PI3K3/AKT, RAS/MEK, TP53, MYC; and we may from this moment regard that some of the aforementioned proteins exist in the Markov chains (TP53, MYC, PIK3).

*JAK/STAT pathway.*

In the instance of the JAK/STAT pathway, there exist in the Markov chain of the chimeric protein some proteins in turn which play roles in this signalling pathway, e.g.: JAK1,

SOCS1, SOCS3, SOS1, GRB2, MYC, PIK3R1, PTPN6, PIK3R2, EGFR, PDGFR (Rawlings, Rosler, & Harrison, 2004); by inhibiting JAK1 activity we could effectively inhibit the signalling pathway, or by means of promoting the transcription of SOCS1 or SOCS3 which effectively inhibit JAK1 through a negative feedback fashion,

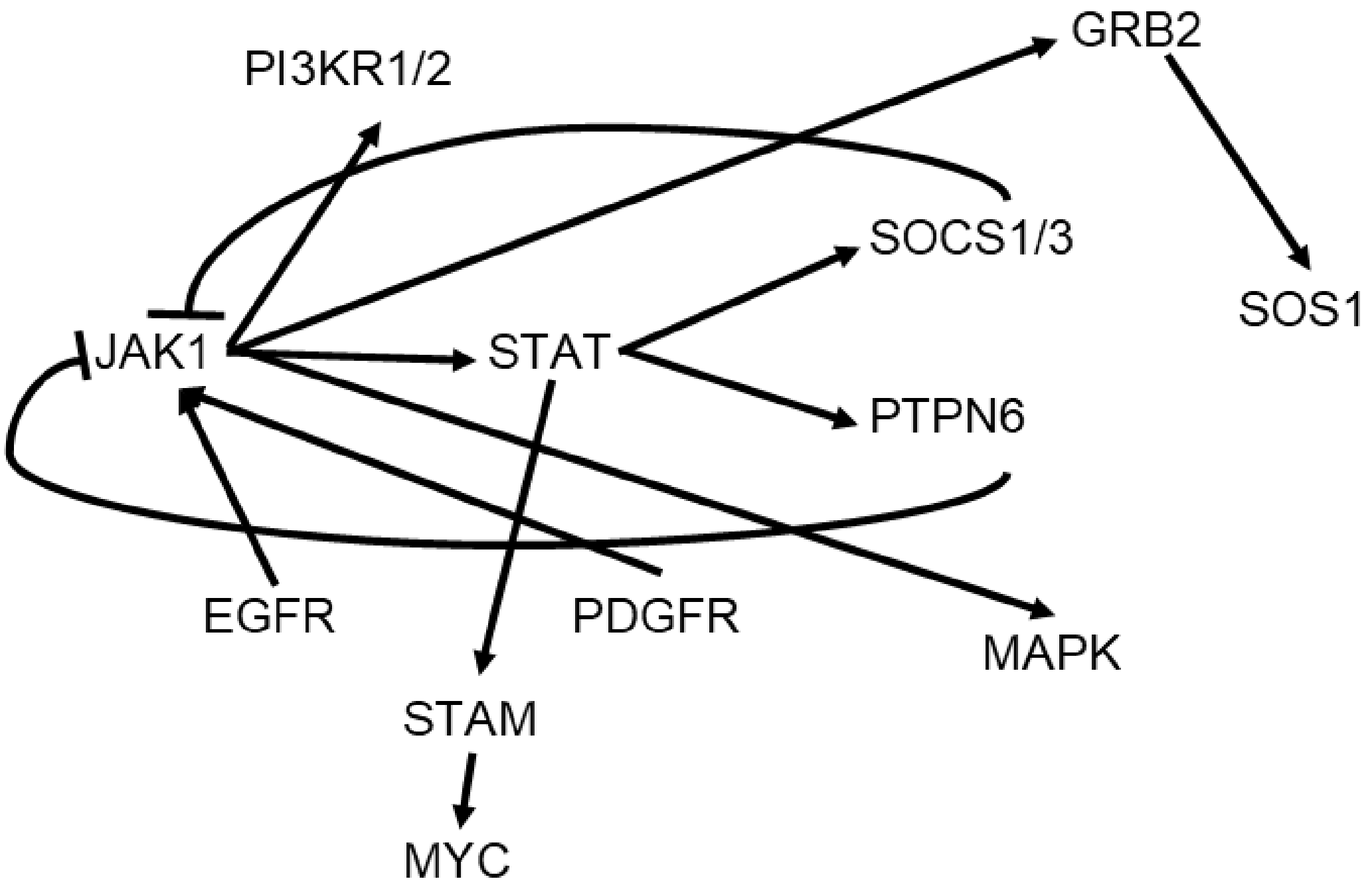

By inhibiting JAK1 would we be targeting $p = 0.00298422$ of the network involved in the chimeric protein, but because of its various interactions, this probability would sum up to $p = 0.11778089$, which would represent a larger inhibition than that represented by the inhibition of BCR or ABL1 on their own; several JAK1 inhibitors have been described, such as ruxolitinib (TyersLab, 2020) and other JAK1-selective inhibitors: filgotinib, GLPG0555, GSK2586184, INCB039110, INCB47986, ABT-494; said inhibitors encounter themselves in clinical phases I, II, IIb and on hold (Menet, Mammoliti, & López-Ramos, 2015). Alternative manners to achieve suppression of JAK1 signaling pathway involve the JAK1-independent promotion of expression of SOCS1/3, the one for SOCS1 has been performed elsewhere through the administration of interferon-gamma (Madonna, et al., 2015). EGFR inhibitors are currently used in different types of cancer (e.g.: non-small cell lung cancer) and are classified in three generations, some of these drugs are: gefitinib, erlotinib, afatinib, dacomitinib, neratinib, osimertinib, rociletinib, olmutinib, ASP8273, nazartinib, avitinib; some of which are either in phases I, II or III and some others have already been approved (Sullivan & Planchard, 2017).

*TP53 pathway.*

TP53, when activated is due to the cellular response to an insult, either in nutrition, low oxygen levels, DNA damage, etc., this activation turns in change in the expression of genes corresponding to cellular senescence and apoptosis, it has been seen that TP53 is activated by ATR and ATM (by means of a third party protein not involved in the main protein network) and inactivated my MDM2 (Kanehisa Laboratories, 2018), the main network spends 5.196799% of its time in this subnetwork and is a feasible target in CML therapy, either by means of TP53 activation, MDM2 inhibition or ATM and ATR activation. As MDM2 inhibitors we could readily use nutlin-3 a small molecule which has been studied as a therapy in some cancers in combination with other drugs, such as 1,25-dihydroxyvitamin D(3), AZD6244, PI-103, MK-0457, perifosine, dasatinib, doxorubicin, platinum compounds (Shen & Maki, Pharmacologic activation of p53 by small-molecule MDM2 antagonists., 2011). ATR activation can be mediated through the administration of hydroxyurea, which has been shown to perform its action by means of increasing replication stress (Landsverk, et al., 2019).

*Beta-catenin pathway.*

It involves the primary condition of whether there are any Wnt ligands that may activate DVL, if the condition is not sufficed, the destruction complex will promote the ubiquitination of CTNNB1 (beta-catenin), this is possible through four proteins: APC, GSK3B, CK1, axin. If the condition is sufficed, the destruction complex will be inhibited by DVL, more specifically, DVL will inhibit the actions of GSK3B and thus of the whole complex; with this inhibition, CTNNB1 is free to roam into the nucleus and promote the expression of genes like MYC and cyclin D1 (Krishnamurthy & Kurzrock, 2018). The inhibition of this pathway would represent the inhibition of 1.918889% of the BCR-ABL1 chimeric protein network and may be done through a series of either inhibitions or activations; we may inhibit the WNT ligands, DVL, or CTNNB1; or we may promote either the expression or the function of APC, GSK3B, CK1, axin. Regarding DVL inhibitors, we have some options, such as: niclosamide, sulindac, NSC668036, J01-017a. Now, with CTNNB1 we have some options, such as the inhibitors of the co-activators: ICG-001, PRI-724; and those of the DNA binding: PKF115-584, PKF118-310 (Krishnamurthy & Kurzrock, 2018).

*Combination therapies.*

In this instance, we propose imatinib or a TKI of second- or third-line of treatment as the cornerstone of therapy, however, because this represents the inhibition of 12.314568% of the network (directly), we propose the use of a branched treatment in order to promote the inhibition of this network's signalling,

| Cornerstone drug. | Branch drug. | | | Inhibition probability. |
|---|---|---|---|---|
| Imatinib. | + | | | $p = 0.12314568$ |
| | JAK/STAT pathway inhibitors.<br>• JAK1 inhibitors.<br>• SOCS1 expression promoters.<br>• EGFR inhibitors. | + | | $p = 0.24092657$ |
| | | TP53 pathway activators.<br>• ATR activators.<br>• MDM2 inhibitors.<br>• TP53 activators. | + | $p = 0.29289456$ |

| | | | |
|---|---|---|---|
| | | CTNNB pathway inhibitors.<br>• DVL inhibitors.<br>• GSK3B activators.<br>• CTNNB1 inhibitors. | $p = 0.31208345$ |

Final remarks.

The prognosis for patients with CML has improved by means of the introduction of imatinib into its treatment, with the advent of different tyrosine kinase inhibitors, resistance in patients could be adverted; however, goals in both the MMR and CCR in patients with CML remain low as well as the diminishment in life expectancy of these patients versus the general population, in this paper we have attempted to provide a Markov chain model to predict the most probable interactions within the networks of the proteins involved in the pathogenesis of CML: BCR, ABL1 and the chimeric protein BCR-ABL1; and by observing the signalling pathways affected by the aberrant behaviour of the chimeric protein we could devise some adjuvant or “branch”-like therapies to the cornerstone one (tyrosine kinase inhibitors) in order to augment the probabilities of patients for a longer life expectancy, quality of life and diminishment in incidence of blastic crises, which severely endanger life.

References.

Addenda.

| Set of molecule/molecule. | Chromosome location. | Comments. |
|---|---|---|
| ⅇ Metabolism. | | |
| HSPA8. | 11q24.1 | It is a member of the heat shock protein family A, functioning as a chaperone binding to newly formed peptides for their appropriate folding; possesses functions in vesicle disassembly (NIH, 2019). BCR-ABL1 has been seen to induce the expression of this protein in CD34(+) CML cells which forms complexes with different gene products induced as well by the fusion protein: CCND1, CDK4 which may contribute to the abnormal proliferation of CML cells (San José-Enériz, et al., 2008). |
| USP15. | 12q14.1 | Ubiquitin specific peptidase 15 (USP15), is a member of the family of deubiquinitating enzymes, it also associates with the COP9 signalosome and shows a role in the deubiquitination of SMAD transcription factors (NIH, 2019). This protein has been reported to be amplified in breast, ovarian cancer and glioblastoma tumours regulating the TGF-beta transcription pathway (Pal & Donato, 2014) |
| HSP90AA1. | 14q32.31 | Heat shock protein pertaining 90 alpha family with chaperone functions in its homodimer form, this functions are performed throughout the use of an auxiliary ATPase activity (NIH, 2019), it has been related to some signalling pathways in cancer cells, and as well it has been seen to regulate the expression of BCR-ABL1, thus could we approach the inhibition of this protein |
| | | to diminish the expression of BCR-ABL1 fusion protein (Shen, et al., 2014). |
| UBC. | 12q24.31 | It is a ubiquitin gene, thus it has been associated with protein degradation, DNA repair, cell cycle regulation, kinase modification, endocytosis and other pathways (NIH, 2019), has it established been that ubiquitination of BCR-ABL1 may be an upcoming therapeutic target in the treatment of CML (Ru, et al., 2016). |
| WDR48. | 3p22.2 | This protein interacts with ubiquitin specific peptidase 1, activating this protein's activity removing ubiquitin from FANCD2 which is in turn ubiquitinated in response to DNA damage (NIH, 2019); WDR48 has been seen in Ph(-) clones of myeloproliferative neoplasms fused to PDGFRB (platelet-derived growth factor receptor beta) (Hidalgo-Curtis, et al., 2009). |
| ERCC3. | 2q14.3 | This gene encodes a DNA helicase ATP-dependent that functions in nucleotide excision repair, it is a subunit of basal transcription factor 2, and it functions in class II transcription. It has been associated to diseases such as: xeroderma pigmentosum B, Cochayne's syndrome, trichothiodystrophy (NIH, 2019). |
| PIK3R1. | 5q13.1 | It encodes a protein which functions as a regulatory subunit to the phosphoinositide-3-kinase which phosphorylates the inositol ring f phosphatidylinositol at 3'. Plays a role in the metabolic actions of insulin, and it has been associated to insulin resistance (NIH, 2019). Moreover, has it been seen that BCR-ABL1 induces TGF-$\beta 1$-activated PI3K signalling which promotes the release of s-KitL and s-ICAM1 |
| | | which in turn augments the presence of tumour stem cells in peripheral circulation (Li, et al., 2015). |
| PDZK1. | 1q21.1 | This gene contains a PDZ-domain, deriving from it its name, being this protein a scaffolding protein. These class of molecules binds to and mediates the subcellular localization of target proteins, in this case it mediates the localization of cell surface proteins, and amongst the proteins whose localization regulates we may find HDL receptor, thus having a role in cholesterol metabolism. It has been associated with metabolic syndrome and resistance of multiple myeloma (NIH, 2019). Recently, it was shown that BCR binds to PDZK1 at the junctional region of epithelial cells, which in turn may affect Ras signalling (Malmberg, et al., 2004). |
| COMMD1. | 2p15 | The encoded protein is a regulator of cooper homeostasis, sodium uptake and NFKB signalling (NIH, 2019). Cellular stress due to BCR-ABL1 has been related to nucleoli disruption and increase in COMMD1 expression (Chen & Stark, 2019). |
| TP53. | 17p13.1 | It is a tumour suppressor protein which may activate transcription, bind to DNA and oligomerize. It responds to cellular stresses, thus being able to induce cycle arrest, apoptosis, senescence, DNA repair, change metabolism. This protein has been related to many forms of cancer, including hereditary ones (NIH, 2019). TP53 has been associated to imatinib in CML cases (Wendel, et al., 2006). |
| MAOA. | XP11.3 | Monoamine oxidase A is a mitochondrial enzyme which catalyzes the deamination of amines. A mutation in this gene could induce Brunner syndrome, and a plethora of psychiatric disorders (NIH, 2019). Inhibiting MAOA could in turn disrupt the blood-brain barrier and then promote a better delivery of imatinib to the CNS compartment (Kast & Focosi, 2010). |
| MCM2. | 3q21.3 | It is a minichromosome maintenance complex component, part of a family of proteins which are highly conserved and are involved in the initiation of eukaryotic genome replication, it regulates the helicase activity of the complex, and it is regulated by protein kinases CDC2 and CDC7 (NIH, 2019). |
| NAGK. | 2p13.3 | N-acetylgucosamine kinase is the major enzyme which recovers amino sugars and catalyzes the conversion of N-acetyl-D-glucosamine to N-acetyl-D-glucosamine 6-phosphate (NIH, 2019). NAGK expression has been inversely correlated with cellular viability (Iorns, et al., 2009). |
| ව Signal processing. | | |
| CSNK2A2. | 16q21 | It is a casein kinase 2 which phosphorylates acidic proteins; involved in several cellular processes: cell cycle, apoptosis, circadian rhythms. An intronic variant could be associated with leukocyte telomere length in South Asian population (NIH, 2019). |
| TRIM25. | 17q22 | Member of the tripartite motif family, it localizes to the cytoplasm and may function as a |
| | | transcription actor; its expression is promoted by oestrogen (NIH, 2019). |
| HCK. | 20q11.21 | Member of Src family of tyrosine kinases, with hemopoiesis functions in the myeloid and B-lymphoid lineages, in the respiratory burst, neutrophil migration and neutrophil degranulation (NIH, 2019). |
| FES. | 15q26.1 | Human cellular counterpart of a feline sarcoma retrovirus protein, with tyrosine kinase activity. Its chromosomal location has been linked to a translocation occurred in patients with acute promyelocytic leukaemia, but also involved in hemopoiesis and cytokine receptor signalling (NIH, 2019). |
| CBL. | 11q23.3 | Proto-oncogene that encodes a RING finger E3 ubiquitin ligase, one of the enzymes necessary for targeting substrates for degradation by proteasome, by means of transfer of ubiquitin from ubiquitin conjugating enzymes. It may also interact with tyrosine-phosphorylated substrates for their degradation by proteasome. When translocated, it could lead to acute myeloid leukaemia (NIH, 2019). |
| CRK. | 17p13.3 | Proto-oncogene that binds to tyrosine-phosphorylated proteins, it possesses two domains with counteracting abilities of transformation (NIH, 2019). |
| KIT. | 4q12 | It is a proto-oncogene, type 3 transmembrane receptor for MGF; mutations in this gene have been associated with various diseases: gastrointestinal stromal tumours, mast cell disease, acute myelogenous leukaemia (NIH, 2019). |
| DOK1. | 2p13.1 | Docking protein 1, is a protein which engages in the downstream signalling pathway of tyrosine kinases, being this protein a scaffold protein which assists in the assembly of signalling complexes (NIH, 2019). |
| ERBB2IP. | 5q12.3 | ERBB2 interacting protein contains a PDZ domain and binds to the unphosphorylated form of ERBB2 protein, regulating it. This protein affects Ras signalling (NIH, 2019). |
| SOS1. | 2p22.1 | It is a guanine nucleotide exchange factor for RAS proteins. By binding to GTP it activates RAS proteins and by its binding to hydrolysed GTP, it inactivates RAS proteins. It’s been associated with gingival fibromatosis 1 and Noonan syndrome type 4 (NIH, 2019). |
| SLA2. | 20q11.23 | Part of the SLAP family of adapter proteins, and may downregulate T and B-cell mediated 9responses (NIH, 2019). |
| WEE1. | 11p15.4 | This is a nuclear protein, a tyrosine kinase catalysing the phosphorylation of CDC2/cyclin B kinase resulting in inhibition of the complex and mediates the coordination of transition between DNA replication and mitosis (NIH, 2019). |
| PTPN6. | 12p13.31 | Member of the protein tyrosine phosphatase family, a signalling molecule involved in cell growth, differentiation, mitotic cycle and oncogenic transformation. This member is expressed in a predilect fashion in hematopoietic cells (NIH, 2019). |
| TULP3. | 12p13.33 | Member of the tubby gene family of bipartite transcription factors. This protein binds to phosphoinositides and functions as a transcription regulator that translocates to the |
| | | nucleus when the phosphoinositide molecule hydrolyzes, thus playing a role in neuronal development and its function (NIH, 2019). |
| • Structure. | | |
| MLLT4. | 6q27 | Multi-domain protein that plays a role in signalling and organization of cell junctions, and it has also been seen as a fusion partner of acute lymphoblastic leukemic gene, which takes part in acute myeloid leukaemias (NIH, 2019). |
| EZR. | 6q25.3 | Ezrin is the encoded protein by this gene, a membrane protein which functions as a tyrosine kinase substrate in microvilli, plays a role in cell surface adhesion, migration and organization. It has been associated with various cancers (NIH, 2019). |
| JPH4. | 14q11.2 | Member of the junctophilin family of transmembrane proteins, involved in junction adhesions (NIH, 2019). |
| RSPH9. | 6p21.1 | Protein component of radial spoke head in cilia and flagella. Mutations in this gene have been associated with primary ciliary dyskinesia 12 (NIH, 2019). |
| SYNC. | 1p35.1 | Member of the intermediate filament family, with a high expression pattern in skeletal and cardiac muscle. It links the dystrophin associated protein complex to desmin filaments (NIH, 2019). |
| • Cell trafficking. | | |
| AP2M1. | 3q27.1 | Subunit of the heterotetrameric coat assembly protein complex 2, this protein is required for the activity of a v-ATPase, which conduces to acidification of endosomes and lysosomes, it has a role in intracellular trafficking and CTLA4 protein function regulation (NIH, 2019). |
| • Cell growth. | | |
| GRB2. | 17q25.1 | This protein binds the epidermal growth factor receptor, and one of its domains binds tyrosine phosphorylated sequences (NIH, 2019). |
| TSG101. | 11p15.1 | Homolog of ubiquitin-conjugating enzymes, it interacts with stathmin, a phosphoprotein implicated in tumorigenesis. It plays a role in cell growth and differentiation and it is a negative growth regulator. It plays a role in genomic stability and cell cycle regulation (NIH, 2019). |
| • Cell proliferation. | | |
| • TP53. | Idem sup. | Idem sup. |

Table 1. BCR-ABL1 fusion protein interacts with various other molecules, a selection of them along with a small description of them is shown above, shedding light about the targets this protein possesses, and the possible roles of these proteins in the development of the disease.

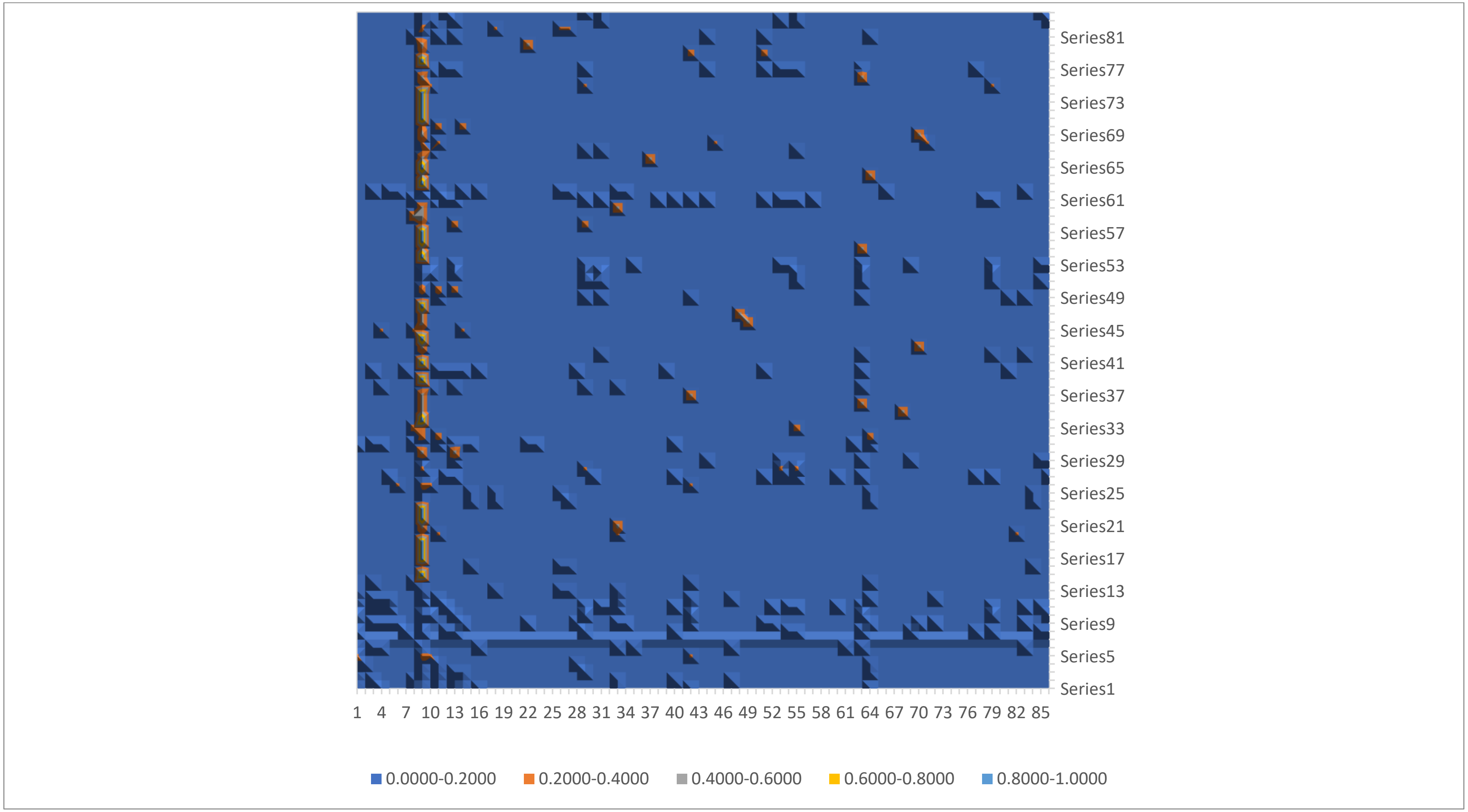

Fig. 1. Graph depicting the stochastic matrix of the BCR protein interactions; in the y-axis we may find the initial state protein and the x-axis represents the molecule towards each of the i-th initial molecule/state will move to in the next step.

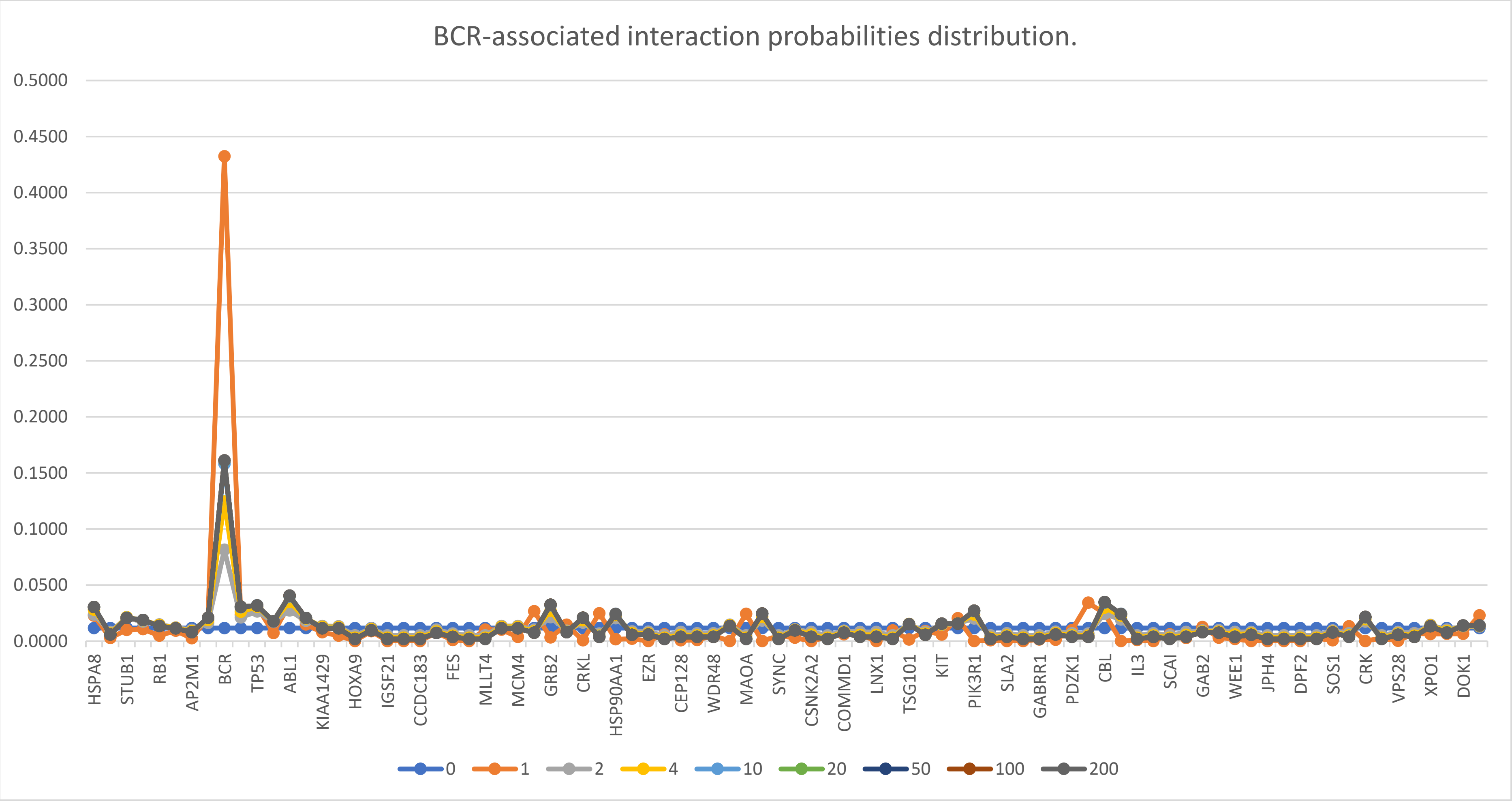

Fig. 2. Stationary probability distributions of the BCR protein interactions, with the x-axis depicting the molecules with which BCR interacts, the y-axis the corresponding probability and the different coloured lines, the values of iteration or steps considered to compute the distribution.

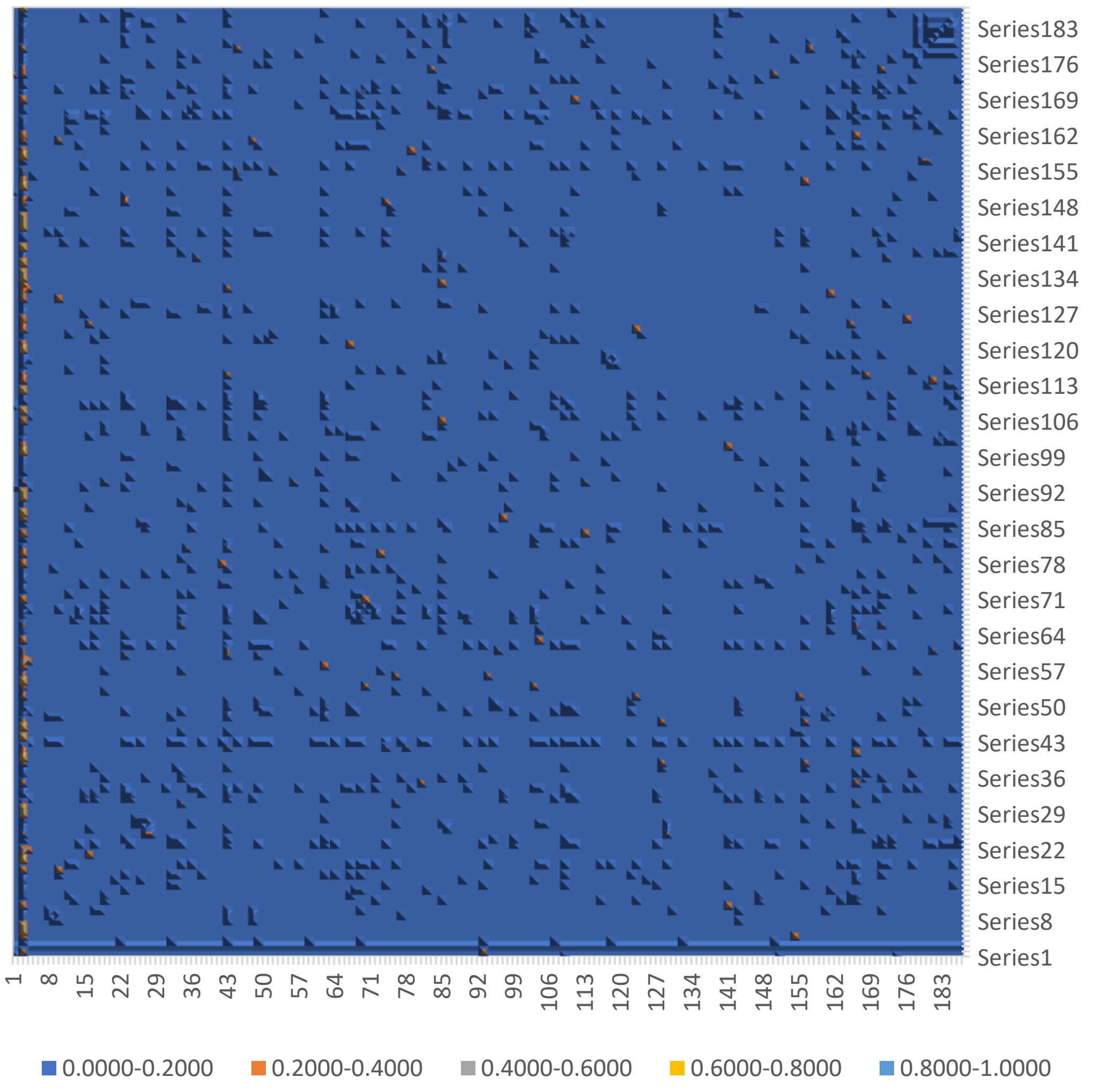

Fig. 3. Graph which represents the stochastic matrix of the ABL1 protein interactions.

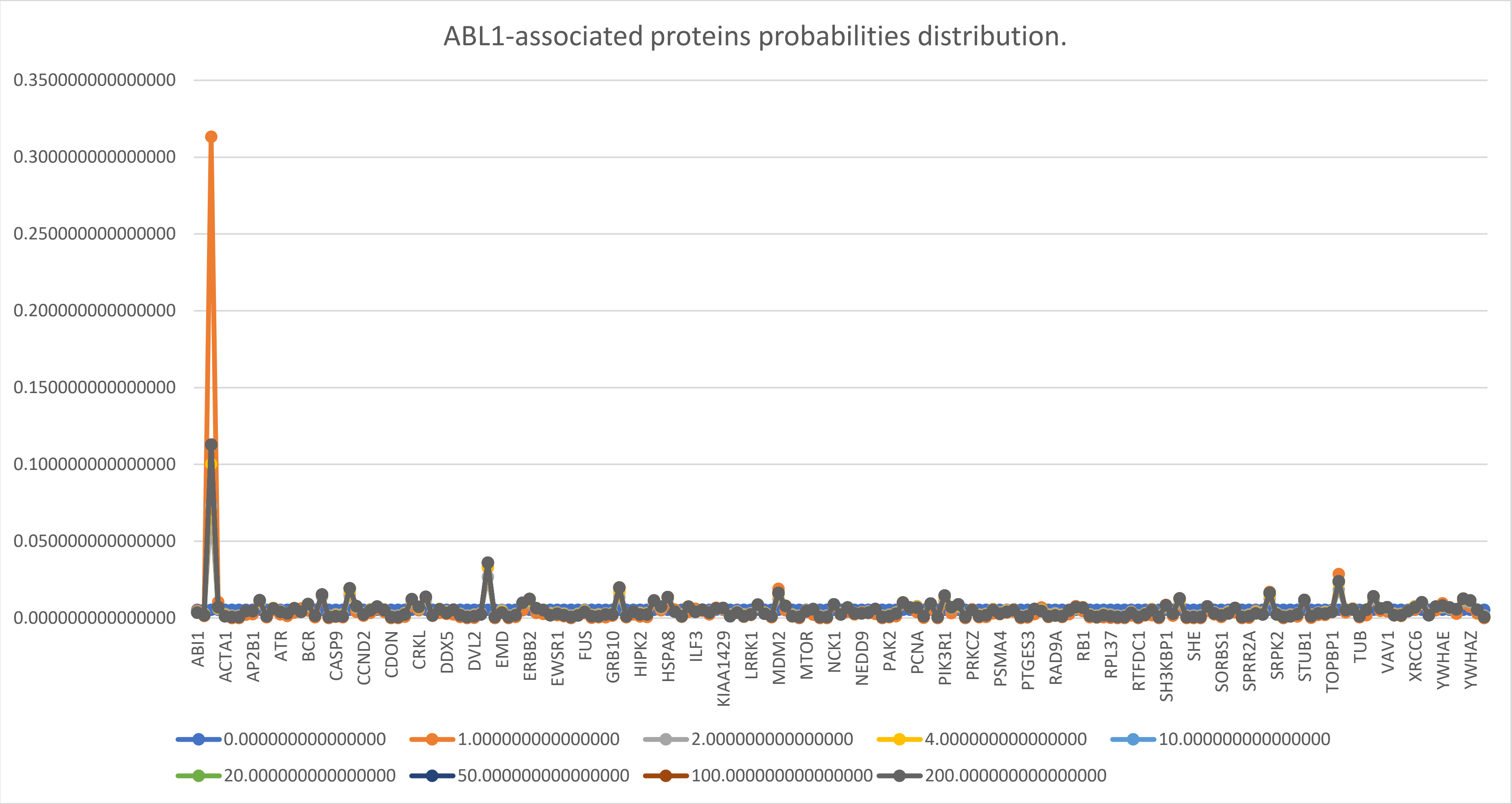

Fig. 4. This graph shows the stationary probabilities distribution of ABL1-associated proteins in different moments of time, with the x-axis representing the various molecules with which ABL1 interacts, y-axis the probabilities and the different lines, the moments at which the probabilities were computed.

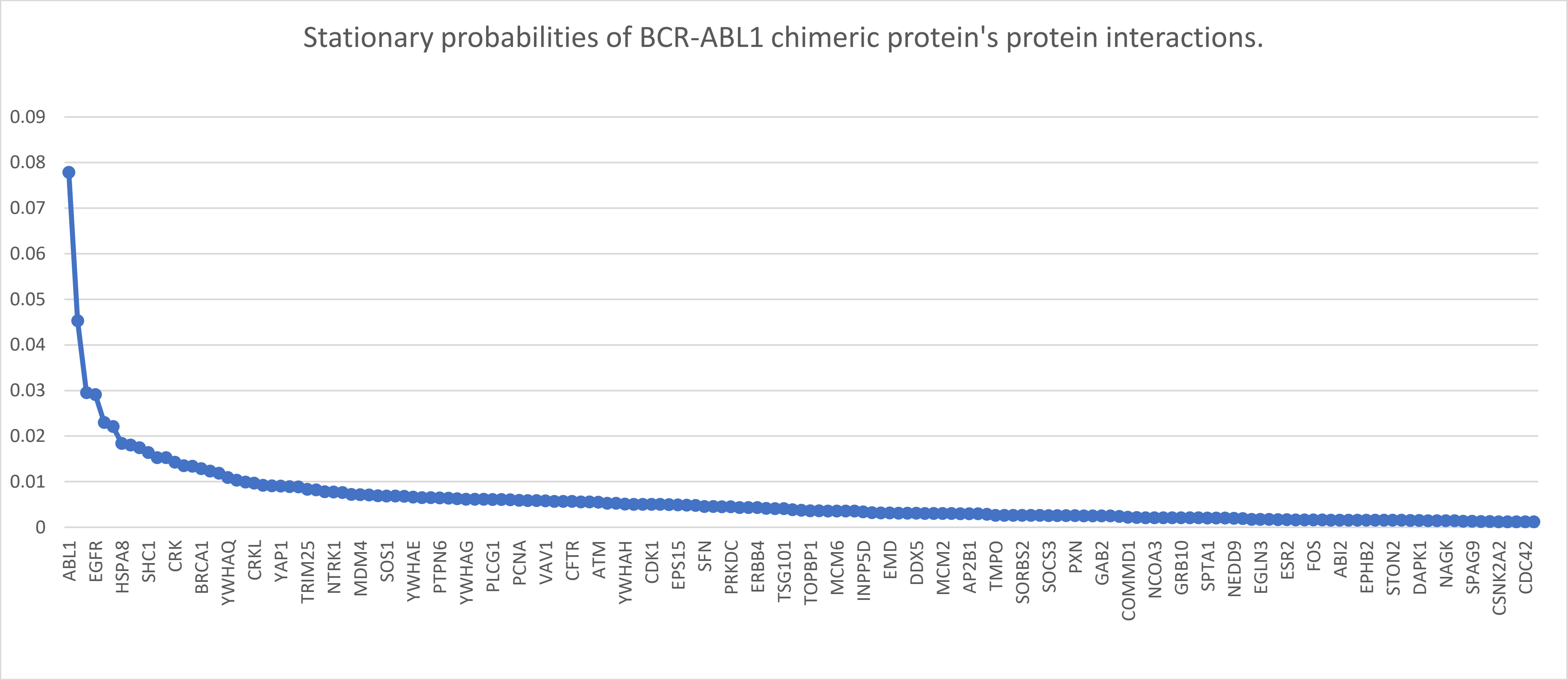

Fig. 5. Stationary probabilities corresponding to those proteins that interact with BCR-ABL1 chimeric protein when n iterations have been performed, here $n = 200$.

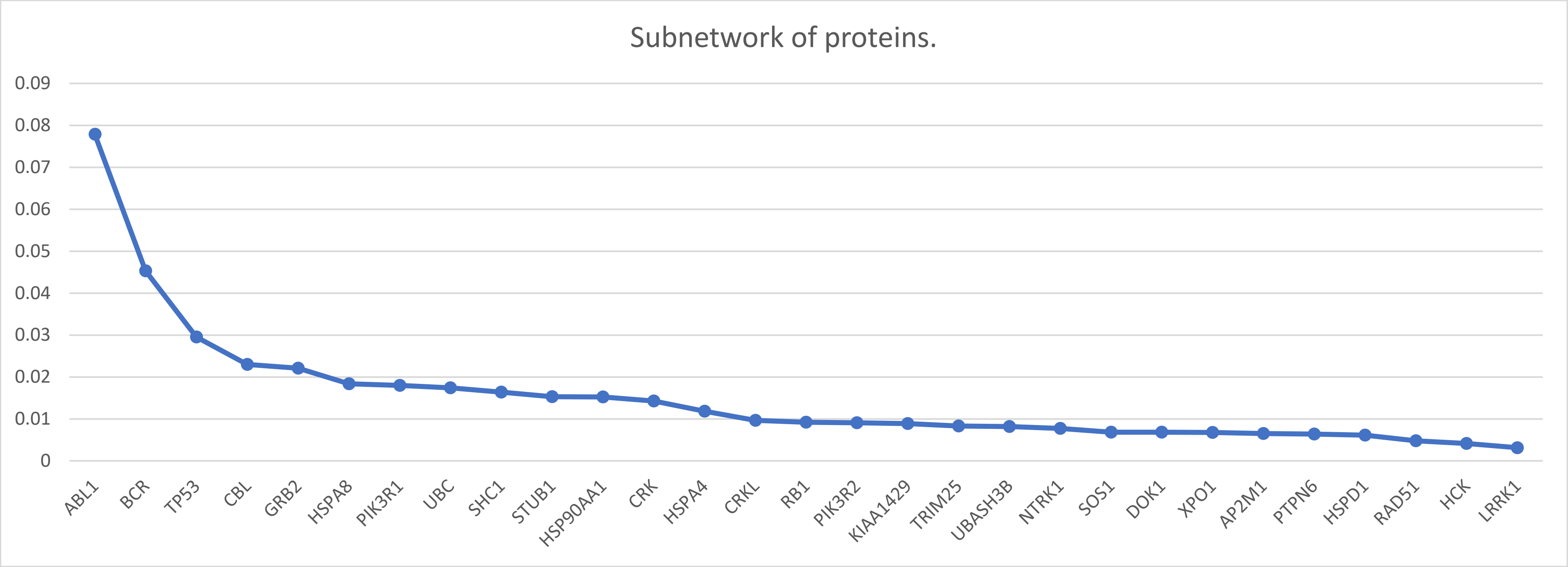

Fig. 6. Graph which shows us the stationary probabilities of those proteins in the Markov chain corresponding to the chimeric protein which are encountered in the intersection of the sets of proteins of BCR and ABL1.

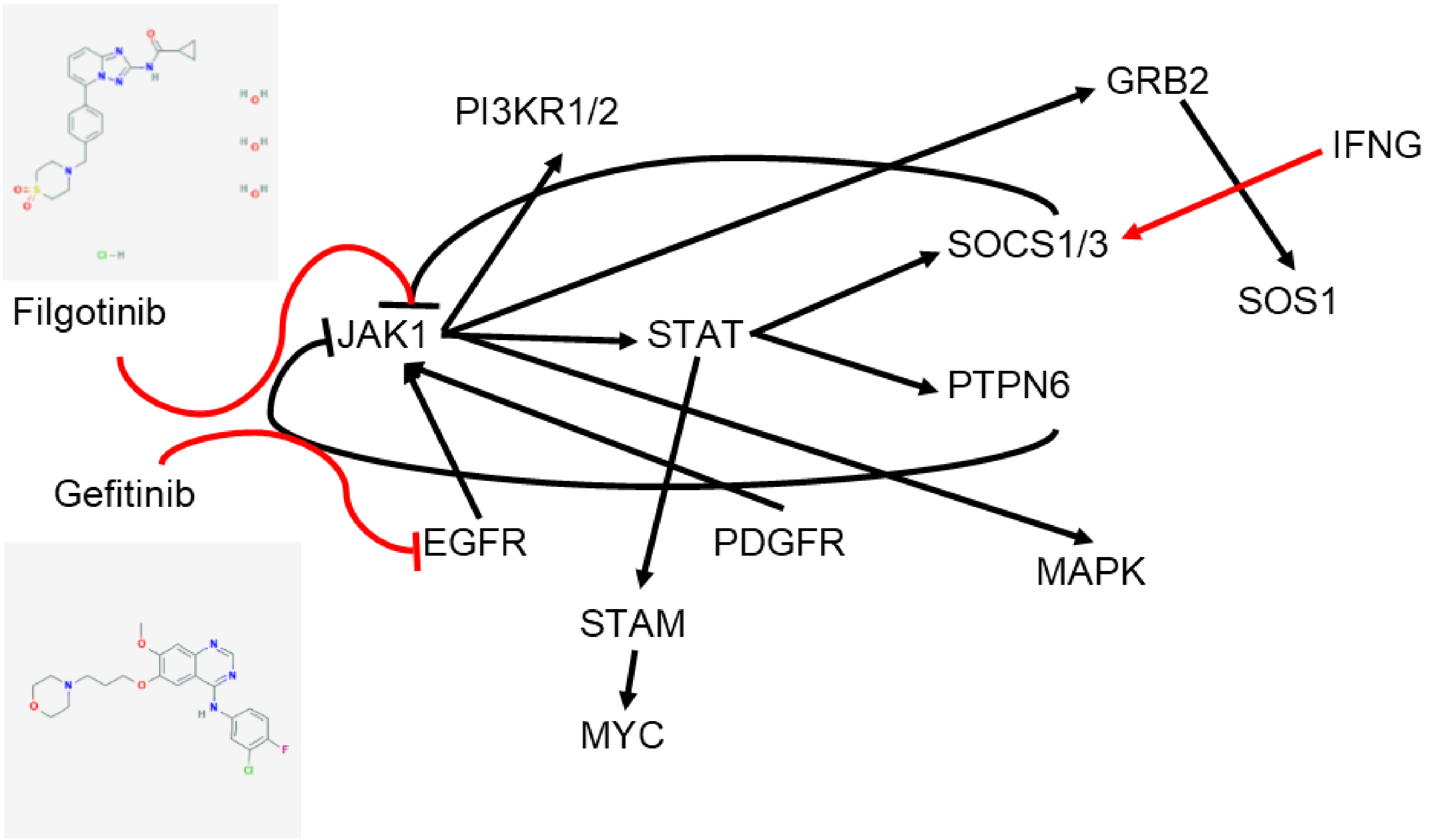

Fig. 7. Network showing the JAK/STAT signalling pathway (Rawlings, Rosler, & Harrison, 2004) and the potential sites of its inhibition: JAK1 inhibition by filgotinib (depicting one of many drugs which could be potentially used) (PubChem, 2020), gefitinib as an EGFR inhibitor (PubChem, 2020) and IFNG as a promoter of SOCS1 expression.